\documentclass[9pt,twocolumn,twoside]{opticajnl}
\journal{opticajournal} % Choose journal (ao,jocn,josaa,josab,ol,optica,pr)

\setboolean{shortarticle}{true}
\usepackage{comment}
\usepackage{outlines} %added by TB.  
\usepackage{hyperref}
\hypersetup{
    colorlinks=true,
    linkcolor=blue,
    filecolor=magenta,      
    urlcolor=cyan,
    pdftitle={Overleaf Example},
    pdfpagemode=FullScreen,
    }
\usepackage{subfig}
\usepackage{caption}
\usepackage{setspace}
\usepackage{xcolor}
\usepackage{float}
\usepackage{subfloat}
\usepackage{soul} %highlighting
\usepackage{listings}  %% https://tex.stackexchange.com/questions/121601/automatically-wrap-the-text-in-verbatim
\usepackage{lineno}

\title{Foundry manufacturing of octave-spanning microcombs}

\author[1,2]{Jizhao Zang}
\author[1,2,*]{Haixin Liu}
\author[1]{Travis C. Briles}
\author[1,2]{Scott B. Papp}

\affil[1]{Time and Frequency Division, National Institute of Standards and Technology, 325 Broadway, Boulder, Colorado 80305, USA}
\affil[2]{Department of Physics, University of Colorado, 440 UCB, Boulder, Colorado 80305, USA}

\affil[*]{Corresponding author: haixin.liu@colorado.edu}

\begin{abstract}

Soliton microcombs provide a chip-based, octave-spanning source for self-referencing and optical metrology. We explore use of a silicon-nitride integrated photonics foundry to manufacture octave-spanning microcombs. By group-velocity dispersion engineering with the waveguide cross-section, we shape the soliton spectrum for dispersive-wave spectral enhancements at the frequencies for f-2f self-referencing. With the optimized waveguide geometry, we control the carrier-envelope offset frequency by adjusting the resonator radius. Moreover, we demonstrate the other considerations for octave microcombs, including models for soliton spectrum design, ultra-broadband resonator external coupling, low-loss edge couplers, and the nonlinear self-interactions of few-cycle solitons. This design process permits highly repeatable creation of soliton microcombs optimized for pump operation less than 100 mW, an electronically detectable offset frequency, and high comb mode power for f-2f detection. However, these design aspects must also be made compatible with the foundry fabrication tolerance of octave microcomb devices. Our experiments highlight the potential to manufacture a single-chip solution for an octave-spanning microcomb, which is the central component of a compact microsystem for optical metrology.

\end{abstract}

\setboolean{displaycopyright}{true}

\begin{document}
\maketitle
%\hl{SP:  please leave the citations alone for now.  Let's get the text in place, first.}
Kerr-microresonator soliton frequency combs, or microcombs, have undergone rapid development because of the insights they provide to nonlinear dynamics and their application possibilities \cite{Kippenberg2018DissipativeMicroresonators,Gaeta2019Photonic-chip-basedCombs}.  
These broadband laser sources are composed of discrete spectral lines with the relation $f_n = n f_{\text{rep}} + f_{\text{ceo}}$, where $n$ is an integer comb mode number, $f_{\text{rep}}$ is the repetition frequency and $f_{\text{ceo}}$ is the carrier-envelope offset frequency. 
Several applications require phase stabilization of $f_{\text{rep}}$ and $f_{\text{ceo}}$, including optical frequency synthesis \cite{spencer2018optical} and measurement with respect to either microwave or optical clock signals, optical clock metrology, and optical frequency division \cite{sun2024integrated}.

Realizing a phase-stabilized microcomb involves generation of an octave-spanning spectrum with sufficient optical power for f-2f self-referencing and phase-locking of the $f_{\text{ceo}}$ and $f_{\text{rep}}$ signals \cite{Briles2018InterlockingSynthesis}. Management of group-velocity dispersion (GVD or dispersion) is critical for all of these tasks. The spectrum of a soliton microcomb is mostly determined by the integrated dispersion, $D_\text{int}=\nu_{\mu}-(\nu_{0}+\text{FSR}\,\mu)$, where $\mu$ is the relative mode number to the pump mode, $\nu_{\mu}$ is the cold cavity resonance frequency and FSR is the free spectral range at the pump \cite{black2022optical}. 
While dark solitons in normal GVD have higher
pump-to-comb conversion efficiency \cite{spektor2024photonic,zang2024laser}, bright solitons in anomalous GVD offer the broadest spectra, characterized by a bandwidth adjustable mostly with the second-order dispersion parameter and coherent emission of dispersive waves (DWs) \cite{Brasch2016PhotonicRadiation} that arise due to higher-order dispersion. We consider dual DW microcombs with a shortwave DW (SWDW) and a longwave DW (LWDW) at higher and lower frequency than the pump laser, respectively. These DWs are designed according to the condition $D_\text{int}$ = 0, and they greatly enhance the comb mode power and bandwidth.
We control octave microcombs mostly by adjusting $D_\text{int}$ with the resonator geometry. Indeed, the SWDW and LWDW are very sensitive to fabrication tolerance. Therefore, an approximate target design exists, however, to date it has only been implemented with carefully controlled, low-volume fabrication, requiring iterative device selection and fabrication or device trimming post-fabrication \cite{moille2022integrated}. For frequency metrology, foundry manufacturing as high-volume fabrication of f-2f microcombs would revolutionize use of the SI second in the optical domain. 

Here, we report a single-chip solution of microcombs for f-2f self-referencing, which we fabricate at volume with a commercial silicon-nitride foundry. We present an optimized microcomb design that supports a harmonic, dual DW spectrum with an electronically accessible $f_{\text{ceo}}$ for the highest efficiency in f-2f detection. To account for the fabrication tolerance of the foundry, a chip carries numerous resonators with a programmed variation in ring width ($RW$) and ring radius ($RR$). We model the target design according to detailed SiN dispersion engineering, optimized microcomb dynamics that enhance DW power for f-2f detection, and wavelength-dependent external resonator coupling across octave span. A new feature of our modeling is an analytical expression for the complex, octave span microcomb. Post-fabrication device screening in concert with our models characterizes the fabrication tolerance to yield optimized microcombs for f-2f detection. We demonstrate that the two DW frequencies and $f_{\text{ceo}}$ can be independently controlled in a predictable fashion by use of resonator geometry. Finally, we discuss post-fabrication manipulation of soliton spectrum through control of the pump laser. This process reveals interesting DW dynamics that can lead to step-changes in the DW frequencies and power. Our work demonstrates a reliable procedure of generating ready-to-use octave-spanning combs through a commercial foundry, and also expands the theory of the DW dynamics in soliton microcombs.

% Here, we report foundry manufacturing of octave soliton microcombs, realizing a single-chip solution to f-2f phase stabilization.  ....This model incorporates dispersion simulations....doesn't rely on computationally intensive LLE simulations

%\hl{this needs something like our goal is to create a single chip solution for f-2f, so we charact} In the experiment, we successfully located a device that is able to generate an octave-spanning microcomb with a  $f_{\text{ceo}}$ below 5.1 GHz. Taking into account the fabrication tolerance, the experimental results agree well with the  prediction of our analytical model and numerical simulation. 

%\begin{outline}
%\1 Here we show...
%\end{outline}

 %we design the microresonators with parameter sweep to enhance their tolerance to fabrication uncertainties. After fabrication, we search on the wafer for the soliton microcombs with the right dispersive wave frequencies and $f_\rm{ceo}$ that can be detected by a low-frequency photodetector. 
 %Fortunately, with commercial foundry, we are able to increase the robustness of device through wafer-scale fabrication and generate ready-to-use octave-spanning combs.  
%% apparently \rm has been deprecated: https://tex.stackexchange.com/questions/151897/always-textrm-never-rm-a-counterexample
%\setlength{\belowcaptionskip}{-5pt}

%\label{section:concept}
 \begin{figure}[t!]
\centering
\includegraphics[width=\linewidth,trim={1.7cm 0.0cm 1.7cm 0.1cm},clip]{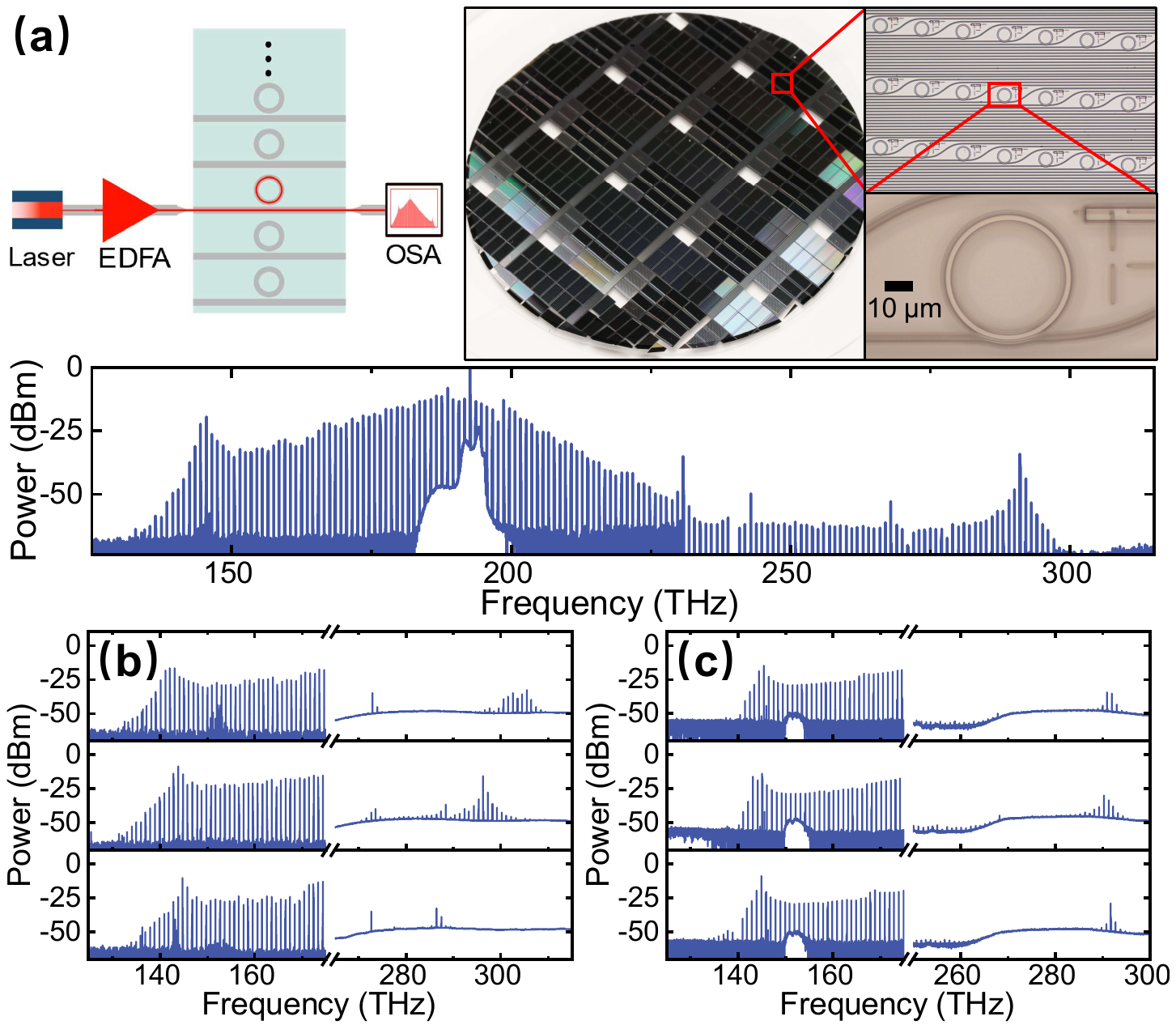}
\caption{\textbf{Key concepts of wafer-scale search for an octave-spanning comb from commercial foundry.} (a) Our chip testing setup (left) and photos of the wafer, a chip, and a resonator (clockwise from middle).  The bottom spectrum shows an octave-spanning microcomb with an on-chip pump power of $(151\pm10)$ mW, $RR$ of 23.475 $\mu$m, and $RW$ of 1690 nm. (b,c) enlargements of DW spectra for different microresonator geometries. The DW locations are sensitive to $RW$ (panel b; $RW$ values: 1670, 1690 and 1710 nm) but less sensitive to $RR$ (panel c; $RR$ values: 23.475, 23.535, 23.58 $\mu$m).  This allows DW control through $RW$ and $f_{\text{ceo}}$ control through $RR$.  
%\hl{old: (b) Zoomed in spectra around dispersive waves with designed RW to be 1670 nm, 1690 nm, and 1710 nm. The designed RR is 23.5 $\mu$m. (c) Zoomed in spectra around dispersive waves with designed RR to be 23.475 $\mu$m, 23.535 $\mu$m, and 23.58 $\mu$m. The nominal RW is 1690 nm.}
}
\label{Fig:concept}
\end{figure}

%\hl{Jizhao, please revise this to focus on the concept of making a chip that has a working microcmb for f-2f on it. Hence, we fabricate devices with a programmed variation in RW and RR.}
We first illustrate in Fig. \ref{Fig:concept} the procedure to obtain an octave-spanning  microcomb through wafer-level and chip-level device selection. The three essential parameters of our devices are layer thickness ($th$), $RW$, and $RR$.  Before fabrication, our analytical model and simulation provided a suggested set of parameters: $th=760$ nm, $RW=1697$ nm, and $RR=23.45$ $\mu $m. The devices were fabricated by Ligentec with a programmed variation in $RW$ and $RR$ around the suggested values. The upper inset of Fig. \ref{Fig:concept} (a) shows photos of our wafer and individual devices (right side), and our experimental setup for soliton generation and characterization (left side). The pump from a C-band continuous-wave (CW) laser is amplified by an EDFA and coupled into a microresonator chip. The chip layer is made of silicon nitride and covered by silicon dioxide top cladding, providing low edge coupling loss ($<3$ dB per edge). We initiate the soliton microcombs by fast sweeping the pump frequency  \cite{Jordan2018,Briles2020GeneratingLasers}, and then finely adjust the detuning to optimize two DWs, monitored by an optical spectrum analyzer (OSA). The lower inset of Fig. \ref{Fig:concept} (a) is a measured spectrum of an octave-spanning microcomb with two DWs at 145.43 THz and 290.92 THz, respectively.  Figure \ref{Fig:concept} (b) and (c) show more details about how we vary $RW$ and $RR$ to tune the DW frequencies and $f_\text{ceo}$ for a workable octave-spanning  microcomb. In Fig. \ref{Fig:concept} (b), we fix $RR$ and sweep $RW$ from 1670 nm to 1710 nm, while in Fig. \ref{Fig:concept} (c),  we fix $RW$ and sweep $RR$ from 23.475 $\mu$m to 23.58 $\mu$m. The measurement results indicate that the DW frequencies can be tuned by $RW$, and they are less sensitive to $RR$. On the other hand, $f_\text{ceo}$ is sensitive to $RR$, which enables us to optimize $f_\text{ceo}$ without changing the DW frequencies. Tuning $RW$ also changes $f_\text{ceo}$ but its influence can be easily counteracted by a relatively wide scan of $RR$ values. With the programmed variation in microresonator geometries, we can identify microcombs with an electronically accessible $f_{\text{ceo}}$ and enhanced DWs at frequencies for f-2f self-referencing.

 %\hl{fix frequencies?}with $f_{ceo}=5.1$ GHz from a device with designed RR and RW to be 23.535 $\mu$m and 1690 nm \footnotemark \footnotetext{\textcolor{red}{am i being dense or is the $|fceo|$ actually 60 GHz??? JZ: 290.92 THz and 145.43 THz  is the data for DWs in Fig. 1(a), fceo=-60 GHz in that case }}. 
 
% \hl{TB: emphasize that DW and fceo can be tuned by RW and RR respectively and it's important that these two controls are orthogonal-ish.  RR variations have only a modest effect on DW positions.  Tuning RW changes fceo somewhat but this easily accounted for by a relatively wide scan of RR values.  Device density is maximized by laying out up to 10 (?) RR per bus. }
 
%\section{Analytic model, Dispersion Design and LLE simulations}
\begin{figure}[htb]
\centering
\includegraphics[width=\linewidth,trim={0cm 0.0cm 0 0.0cm},clip]{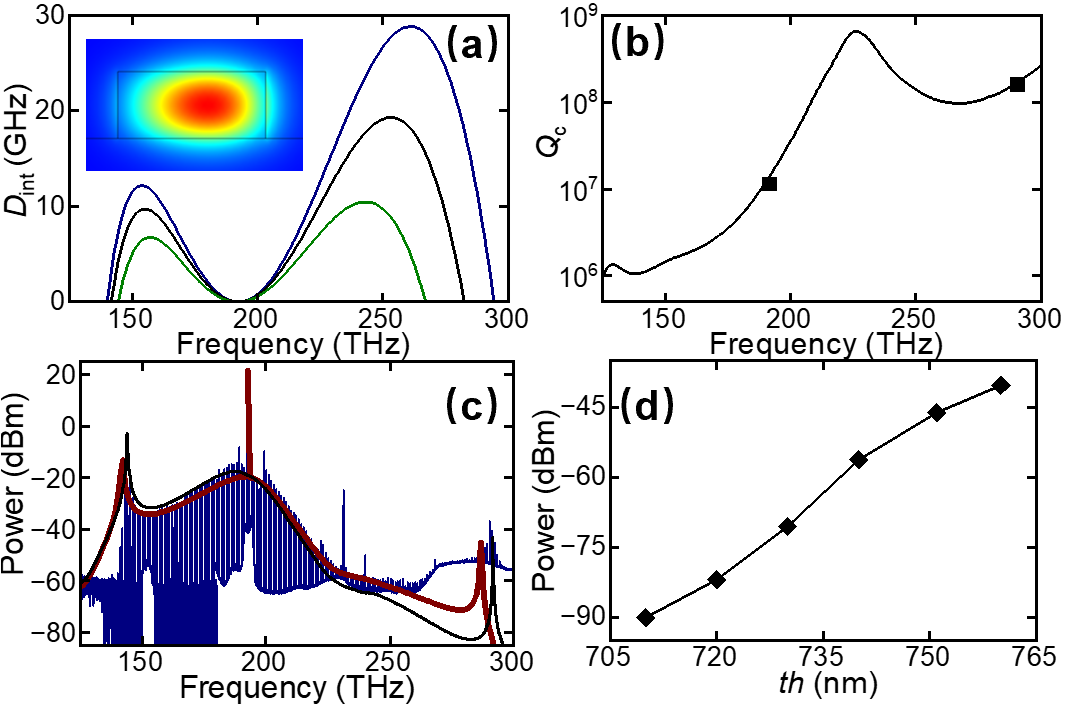} \caption{\textbf{Analytic model and simulation of comb spectrum.} (a) The simulated $D_{\textrm{int}}$ curves with $th$ = 760 nm, $RR$ = 23.45 $\mu$m and $RW$ = 1670 nm (blue), 1697 nm (black), and 1730 nm (green). The inset shows a typical mode profile. %The diagram on the left side shows a typical mode profile in our ring. 
(b) Simulated (line) and measured (squares) $Q_{\textrm{c}}$ vs frequency for $RR$ = 23.535 $\mu$m, $WGW$ = 500 nm, $g_c$ = 950 nm, and $L$ = 17 $\mu$m. %The black line plots the simulation with th = 751 nm and RW = 1659 nm, The black squares are measured $Q_{\textrm{c}}$ on a device with th = 760 nm and RW = 1690 nm.%
%Frequency dependence of $Q_{\textrm{c}}$ with The black curve is the simulated $Q_{\textrm{c}}$ with th$??$nm, RW=$??$ nm. The black squares are the measured $Q_{\textrm{c}}$ with th and RW designed to be ?? nm and 1690 nm. 
 (c) A comparison among measured comb spectra (blue), prediction from our analytic approximation (black) and LLE simulation (red) with  $Q_{\textrm{i}}$ = 1.4M, $F^2$ = 37.2, $\alpha$ = 33.  (d) SWDW power vs $th$, predicted by our analytic model with $F^2$ = 37.2 ($P_{\textrm{in}}$ = 150 mW) and $\alpha$ = 33. For each $th$ value, we adjust $RW$ to ensure that the two DWs span an octave.}
%% RR is fixed at 23.45 $\mu$m.%
%comments: (a) th=760nm, RW=1670,1690,1710nm. (b) th=760nm, RW=1690nm (c) th=760nm RW=1690nm
\label{Fig:model}
\end{figure}

Next, we will discuss the steps to generate a pre-fabrication design. We first regard $th$ as a variable. To ensure an octave span, for each $th$ value, we find the proper $RW$ by using COMSOL software to calculate $D_{\textrm{int}}$ since DWs appear at $D_{\textrm{int}}\approx0$. Figure \ref{Fig:model} (a) shows simulated $D_{\textrm{int}}$ with varied $RW$. The image on the left side shows a typical mode field distribution. The blue, black, and green curves plot the $D_{\textrm{int}}$ with $RW$ of $1670$nm, $1697$nm, and $1730$nm, respectively. For the black curve, two solutions of $D_{\textrm{int}}=0$ span an octave, indicating that $1697$ nm is the proper $RW$ for this $th$ value. 

After $RW$ is determined, we design an optimal coupler that maximizes the coupling rate, denoted by $\kappa_{\textrm{c}\mu}$, at the SWDW, since $\kappa_{\textrm{c}\mu}$ is usually smaller at short wavelength. The coupling is determined by three parameters: the bus waveguide width ($WGW$), the gap between the bus waveguide and the ring, $g_c$, and the pulley length, $L$. We use Lumerical FDTD solver to simulate coupling quality factor, defined by $Q_{\textrm{c}}=2\pi\nu_{\mu}/\kappa_{\textrm{c}\mu}$, and find the best values of $WGW$, $g_\text{c}$, and $L$. The curve in Fig. \ref{Fig:model} (b) shows the simulated $Q_{\textrm{c}}$ with $th$ = 751 nm, $RR$ = 23.535 $\mu$m, $RW$ = 1659 nm, $WGW$ = 500 nm, $g_c$ = 950 nm, and $L$ = 17 $\mu$m. The black squares are measured $Q_{\textrm{c}}$, which agrees well with the simulation considering the fabrication uncertainty. 

We now present a quasi-analytic model for solitons generated in resonators with arbitrary $D_{\text{int}}$ profiles and use it to quantitatively predict the spectral 
mode distribution of optical power, $P_{\mu}$, given the device parameters and some estimation about the pump laser. This model is critical for obtaining sufficient DW power for f-2f self-referencing. Here, we define a mode-dependent conversion efficiency $\textrm{CE}_{\mu}=P_{\mu}/P_{\textrm{in}}$, where $P_{\textrm{in}}$ is the input pump power ($\approx150$ mW in our experiment). $P_{\mu}$ can be estimated from $\text{CE}_{\mu}$ which is related to both the field inside the resonator and the coupling. The intracavity dynamics of Kerr microresonator is governed by the normalized LLE \cite{liu2024threshold}.  
\begin{equation}
\frac{\partial E_{\mu}}{\partial t}=-\left(l_{\mu}+i\left(\alpha+D_\mu\right)\right)E_{\mu}+F\delta_{\mu,0}+i\sum_{\mu_1,\mu_2}E_{\mu_1}E_{\mu_2}E_{\left(\mu_1+\mu_2-\mu\right)}^*.
\label{LLE}
\end{equation}
Here, the intracavity field is decomposed into different modes $E_{\mu}$. We account for frequency dependent loss in terms of the ratio of loss rate between mode $\mu$ and the pump ($\mu=0)$, $l_{\mu} =(\kappa_{\textrm{i}\mu}+\kappa_{\textrm{c}\mu})/(\kappa_{\textrm{i}0}+\kappa_{\textrm{c}0})$, where $\kappa_{\textrm{i}\mu}$ is the intrinsic loss rate. $\alpha$ is the detuning of the pump laser, normalized by the halfwidth of pump mode $(\kappa_{\textrm{i}0}+\kappa_{\textrm{c}0})/(4\pi)$. $D_{\mu} = 4\pi D_{\textrm{int}}/(\kappa_{\textrm{i}0}+\kappa_{\textrm{c}0})$ is the normalized $D_{\text{int}}$. $F$ is the normalized driving force and related to $P_{\text{in}}$ and threshold power $P_{\text{thre}}$ by $F^2 =P_{\text{in}}/P_{\text{thre}}$.

The relation between coupling and $\text{CE}_{\mu}$ is characterized by $r_{\mu} = 2K_{\mu}/(K_{\mu}+1)$ where $K_{\mu}=\kappa_{\textrm{c}\mu}/\kappa_{\textrm{i}\mu}$ is the coupling coefficient of mode $\mu$, and $\textrm{CE}_{\mu}$ has the following form \cite{liu2024threshold}:
\begin{equation}
\textrm{CE}_{\mu} = \frac{|E_{\mu}|^2l_{\mu}r_{\mu}}{F^2/r_{0}}\times\frac{\omega_{\mu}}{\omega_{0}},
\label{CE}
\end{equation}
where $\omega_{\mu}$ is the output angular frequency of mode $\mu$. As discussed above, $\kappa_{\textrm{c}\mu}$ can be calculated from $Q_{\textrm{c}}$. While $Q_{\textrm{c}}$ varies greatly over frequency, the intrinsic quality factor $Q_{\textrm{i}}=2\pi\nu_{\mu}/\kappa_{\textrm{i}\mu}$ has much smaller variation. We assumed $Q_{\textrm{i}}$ to be a constant measured in previous fabrication. Then, $l_{\mu}$ and $r_{\mu}$ can be calculated from $Q_{\textrm{c}}$. 
% we now present an extension of the analytic model for solitons in resonators with purely quadratic dispersion and ...
% our model allows us to uncover previously unknown relations that quantitatively relate the DW power to the fine details of the dispersion landscape...
% is distributed depending on the  the distribution of optical power on highlight this model the uncover previously unknown relations between that allows Should basically say that we extend the simple $Sech^2$ analytical description of soliton spectrum to higher order dispersion, critical for predicting the spectrum of dispersive wave enhanced octave spanning combs...}  
Equation \ref{CE} suggests that we can calculate $\text{CE}_{\mu}$ if we know $E_{\mu}$ and $F$. In resonators with low loss and purely quadratic dispersion, $D_{\mu} = d_2\,\mu^2/2$, the soliton spectra is approximated by \cite{herr2014temporal}:
\begin{equation}
    E_{\mu}^{(0)}=\frac{\sqrt{d_2}}{2}\textrm{Sech}\left[\frac{\pi}{2}\sqrt{\frac{d_2}{2\alpha}}\mu\right].
\label{0th}
\end{equation}
For non-quadratic dispersion, currently no model can approximate the soliton spectra. Here, we provide two extra orders of correction for $E_{\mu}$ which can characterize the spectrum around DWs through perturbation method. To distinguish Eqn. \ref{0th} from our correction, we denote it with a superscript (0).

Our model assumes that $D_\mu\approx d_2\,\mu^2/2$ for small $\mu$ but  diverges from quadratic shape at large $\mu$. For a steady comb, $\partial E_{\mu}/\partial t=i\lambda E_{\mu}$, where $\lambda$ is the time rate of change of the phase and equals the difference between $f_{\text{rep}}$ and FSR. For a $D_{\text{int}}$ with even parity, $\lambda$ = 0. We assume $\lambda$ = 0 in our model but will discuss later that nonzero $\lambda$ will result in DW switching phenomenon. As a result, for nonzero $\mu$, we rewrite Eqn. \ref{LLE} into:

\begin{equation}
E_{\mu}=\frac{\sum_{\mu_1,\mu_2}E_{\mu_1}E_{\mu_2}E_{\left(\mu_1+\mu_2-\mu\right)}^*}{-i l_{\mu}+\alpha+D_\mu}.
\label{steady}
\end{equation}
With this equation, we can calculate our first order solution $E_{\mu}^{(1)}$ by replacing $E_{\mu}$ on the right side with $E_{\mu}^{(0)}$.
%\begin{equation}
%E_{\mu}^{(1)}=\frac{\sum_{\mu_1,\mu_2}E_{\mu_1}^{(0)}E_{\mu_2}^{(0)}E_{\left(\mu_1+\mu_2-\mu\right)}^{(0)*}}{-i l_{\mu}+\alpha+D_\mu}.
%\end{equation}
Instead of directly doing the summation, a more clever way is to notice that $E_{\mu}^{(0)}$ is a good approximation when $D_{\mu}$ is exactly $d_2\,\mu^2/2$. So, according to Eqn. \ref{steady},  $\sum_{\mu_1,\mu_2}E_{\mu_1}^{(0)}E_{\mu_2}^{(0)}E_{\left(\mu_1+\mu_2-\mu\right)}^{(0)*}\approx(-i l_{\mu}+\alpha+d_2\,\mu^2/2)E_{\mu}^{(0)}$, and $E_{\mu}^{(1)}$ has the following expression:
\begin{equation}
E_{\mu}^{(1)} = c_{\mu}^{(1)} E_{\mu}^{(0)},\quad 
c_{\mu}^{(1)} = \frac{-i l_{\mu}+\alpha+d_2\,\mu^2/2}{-i l_{\mu}+\alpha+D_\mu},
\label{1st}
\end{equation}
where $c_{\mu}^{(1)}$ is the first order correction. This expression already gives insight into comb power distribution across a dispersion profile. A striking feature of DWs is that $P_{\mu}$ has local maximum at modes where $D_{int} \approx 0$. This is explained by Eqn. \ref{1st}: for most modes, $D_{\mu}\gg\alpha$, $l_{\mu}$, and $|c_{\mu}^{(1)}|^2$ is relatively small, but when $D_{\mu}$ decreases to $-\alpha$, the lineshape of $|c_{\mu}^{(1)}|^2$ becomes a Lorentzian with peak value $\approx\left(d_2\,\mu^2/(2l_{\mu})\right)^2$.

A more accurate spectrum needs the second order correction:
\begin{equation}
E_{\mu}^{(2)}=E_{\mu}^{0}\frac{\sum_{\mu_1,\mu_3}c_{\mu_1}^{(1)}c_{\mu_3-\mu_1}^{(1)}c_{\left(\mu_3-\mu\right)}^{(1)*}E_{\mu_1}^{(0)}E_{\mu_3-\mu_1}^{(0)}E_{\left(\mu_3-\mu\right)}^{(0)*}/E_{\mu}^{(0)}}{-i l_{\mu}+\alpha+D_\mu}.
\end{equation}
Note that $\textrm{Sech}[x]\leqslant2\,\textrm{Exp}[-|x|]$ and $|\mu_1|+|\mu_3-\mu_1|+|\mu_3-\mu|\geqslant|\mu|$. $E_{\mu_1}^{(0)}E_{\mu_3-\mu_1}^{(0)}E_{\left(\mu_3-\mu\right)}^{(0)*}/E_{\mu}^{(0)}$ vanishes exponentially except for the case where $\mu\leqslant\mu_3\leqslant\mu_1\leqslant0$ or $\mu\geqslant\mu_3\geqslant\mu_1\geqslant0$, and $E_{\mu_1}^{(0)}E_{\mu_3-\mu_1}^{(0)}E_{\left(\mu_3-\mu\right)}^{(0)*}/E_{\mu}^{(0)}\approx d_2$. Then $E_{\mu}^{(2)}$ can be expressed as:
\begin{equation}
E_{\mu}^{(2)} = c_{\mu}^{(2)} c_{\mu}^{(1)} E_{\mu}^{(0)}, 
c_{\mu}^{(2)}=\frac{-i l_{\mu}+\alpha+d_2\sum_{\mu_1,\mu_3}c_{\mu_1}^{(1)}c_{\mu_3-\mu_1}^{(1)}c_{\left(\mu_3-\mu\right)}^{(1)*}}{-i l_{\mu}+\alpha+d_2\,(|\mu|+1)(|\mu|+2)/2},
\label{2nd}
\end{equation}

\noindent where the summation is done over $\mu\leqslant\mu_3\leqslant\mu_1\leqslant0$ or $\mu\geqslant\mu_3\geqslant\mu_1\geqslant0$. In $c_{\mu}^{(2)}$, we retain $-i l_{\mu}+\alpha$ in the numerator, and replace $\mu^2/2$ with $(|\mu|+1)(|\mu|+2)/2$. 
This modification makes $c_{\mu}^{(2)}$ converge to 1 when $c_{\mu}^{(1)}=1$.

%\footnotemark
%\footnotetext{\textcolor{red}{Also I don't understand the stuff about replacing $\mu^2/2$ at all...   Somehow Eq. \ref{2nd} appears to have lost all higher order dispersion information which seems weird because HOD information is in the first order model.  I guess that information is effectively retained because $E_{\mu}$ depends on first and second order corrections? }}

Using  Eqn. \ref{CE} and \ref{2nd}, we can calculate the spectrum of output comb with given values of  $D_{\mu}$, $l_{\mu}$, $r_{\mu}$, $F$, $\alpha$ and $P_{\textrm{in}}$. Equation \ref{0th} indicates that larger $\alpha$ leads to more power, as observed in experiment as well. While absent from Eqn. \ref{2nd}, $F^2$ affects $P_{\mu}$ indirectly by altering the soliton existence range, for which the upper bound of $\alpha$ is $\pi^2F^2/8$ \cite{herr2014temporal}. During the design, we set $F$ based on our laser power and the measured $P_{\text{thre}}$, and perform numerical simulation with Matlab, based on Eqn. \ref{LLE} directly, to estimate the maximum $\alpha$ and the comb spectrum more accurately. Figure \ref{Fig:model} (c) is a comparison among our analytic approximation (black curve), LLE simualtion (red curve), and measured spectrum (blue trace) with nominal $RW$ and $th$ to be 1690 nm and 760 nm, respectively. For the analytic approximation and numerical simulation, we use the same device parameters as in Fig. \ref{Fig:model} (b) and $Q_\textrm{i}$ is 1.4M. Considering the fabrication uncertainty, the analytic, simulated and experimental spectra agree well.
 %Considering the fabrication tolerance, RW = 1659nm and th =751 nm is used in the analytic approximation and numerical simulation for comparison.  
 %$F=6.1$ which is calculated from $P_{\textrm{thre}}=150\pm??$mW and $P_{\textrm{in}}=4.0\pm??$mW, measured in experiment. We set $\alpha=33$ which is the maximum $\alpha$ that can support soliton in our simulation. 

The last parameter to optimize is $th$. We calculate the SWDW power versus $th$ using our analytic model, shown in Fig. \ref{Fig:model}(d). Here, we set $RR$ = 23.45 $\mu$m, $P_{\textrm{in}}$ = 150 mW, $F^2$ = 37.2, and $\alpha$ = 33. $RW$ is adjusted to keep an octave span of DWs. It shows that $th$ = 760 nm will yield the highest SWDW power.

%\section{GVD, SDW/LDW and $f_{\text{ceo}}$ tuning}

%\hl{TB: Now that it's sunk in more, I'm thinking that maybe this section should come after Fig. 1???  The GVD tuning stuff relates directly to Fig. 1 beacuse they are both only about tuning comb frequencies and the model figure + Fig 4 are both really about large DW power.  It would also be a bit easier to write the text about how to use laser control for intense DWs}

%While we have introduced our model in last section, in this section we will use specific values to investigate the sensitivity of octave comb characters to the ring parameters, so that we can sweep RW amd RR on the wafer to cover the fabrication uncertainty.
\begin{figure}[htb]
\centering
\includegraphics[width=\linewidth,trim={0cm 0.0cm 0 0.0cm},clip]{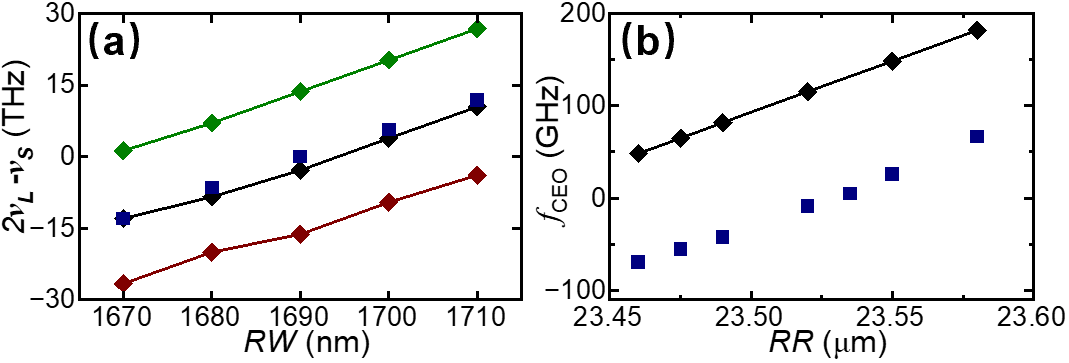}
\caption{\textbf{ Tuning of DW frequencies and $f_{\text{ceo}}$}  (a) $2\nu_{L}-\nu_{S}$ vs $RW$ for $th$ = 754 nm (green), 760 nm (black) and 766 nm (green). The blue squares show the measured data for $th$ = 760 nm. (b) Simulated (black curve with diamonds) and measured (blue squares) $f_{\text{ceo}}$ as a function of $RR$. }
%%(b) th=754,760,766nm RW=1690nm (c)th=760nm RW=1690nm
\label{Fig:data}
\end{figure}

With the suggested device parameters, we next determine their sweep ranges based on the sensitivity of DWs and $f_{\text{ceo}}$. To characterize the harmonic mismatch between two DWs, we define the quantity $2\nu_{\text{L}}-\nu_{\text{S}}$, where $\nu_{\text{L}}$ and $\nu_{\text{S}}$ are LWDW and SWDW frequencies, and plot it as a function of $RW$ and $th$ with a fixed $RR =23.5$  $\mu m$ (Fig. \ref{Fig:data} (a)). The green, black, and red curves with diamonds show the theoretical values with $th$ = 754 nm, 760 nm and 766 nm, respectively. The blue squares are calculated from measured $\nu_{\text{L}}$ and $\nu_{\text{S}}$. The plot shows that 1 nm uncertainty in $th$ requires 5 nm sweep in $RW$ to compensate.

\begin{comment}
As mentioned in Section \ref{section:concept}, $f_{\text{ceo}}$ can be well controlled by tuning $RR$. We can calculate $f_{\text{ceo}}$ by using the formula %\footnotemark \footnotetext{I think you mean the equation I've written in Eq. \ref{eq:fceoDEFN} correct? },
%\begin{equation}
%f_{\text{ceo}} = \left\{\nu_{0}/\text{FSR}\right\}\,\textrm{FSR}
%\end{equation}
%\textcolor{red}{?????} 
%means the fraction part, and $\nu_{0}$ and FSR can be obtained from dispersion simulation. 

\begin{equation}   
f_{\text{ceo}} = f_p - N' \times f_{\text{rep}}
\label{eq:fceoDEFN}
\end{equation}

\noindent where $f_p$ is the pump frequency and $N'$ is the integer part of the ratio $\nu_0/f_{\text{rep}}$. In practice, $f_p$ is not exactly equal to $\nu_0$ due to detuning and $f_{\text{rep}}$ is slightly different from the resonator free spectral range due to nonlinear effects in the soliton (and thermal...). However, these differences can be ignored compared to uncertainty from the fabrication. So, we replace $\nu_p$ and $f_{\text{rep}}$ with $\nu_0$ and FSR in theoretical prediction:
\begin{equation}
    f_{\text{ceo}} \approx \nu_0 - N' \times \textrm{FSR}.
    \label{eq:fceoGVDdefn}
\end{equation}
\end{comment}

Figure \ref{Fig:data} (b) plots the theorectical (black curve with diamonds) and measured (black squares) $f_{\text{ceo}}$ over $RR$ with $RW =1690$ nm and $th =760$ nm. The lowest measured $f_{\text{ceo}}$ is 5.1 GHz with $RR = 23.535$ $\mu$m and the corresponding spectrum is shown in Fig. \ref{Fig:model} (c). Despite an overall shift between prediction and experimental result due to fabrication tolerance, the similar linear dependence and slope over $RR$ highlight the accuracy of our prediction. Since $f_{\text{ceo}}$ is always less than FSR ($\sim$ 1 THz ), a sweep range of 0.8 $\mu$m in $RR$ will always compensate the fabrication uncertainties and enable an  electrically accessible  $f_{\text{ceo}}$.

%\section{Dependence on pump laser of comb spectrum}

\begin{figure}[htb]
\centering
\includegraphics[width=\linewidth,trim={0cm 0.0cm 0 0.0cm},clip]{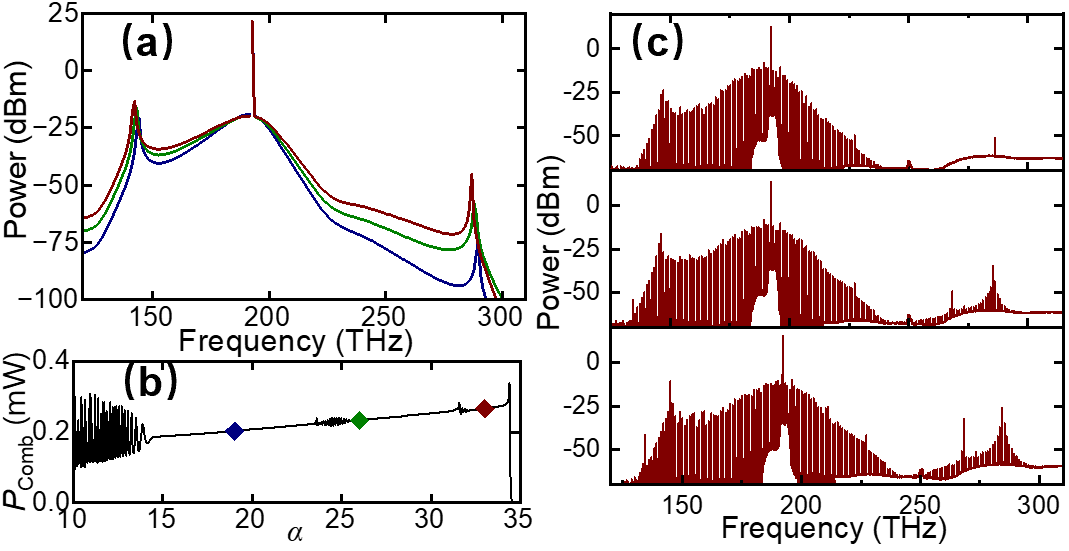}
\caption{\textbf{Dependence of comb spectrum on pump laser.} (a) Simulated soliton spectra with $F^2=37.2$, $\alpha$ = 19 (blue), 26 (green), and 33 (red) using resonator parameters from Fig. \ref{Fig:model} (c) except $\alpha$. ($\mu_{\text{L}}$, $\mu_{\text{S}}$) are (-51,102) (blue), 
 (-52,100) (green), and (-53,99) (red). (b) Simulated comb power as a function of $\alpha$. Each colored diamond corresponds to the soliton spectrum with the same color in (a). (c) Measured soliton spectra with the same devices but different $P_{\text{in}}$: 74 mW (top), 89 mW (middle), and 126 mW (bottom).}
%% (a)(b) F=6 th=760nm RW=1690nm
\label{Fig:pump}
\end{figure}

Besides device design, we can also use post-fabrication methods to manipulate soliton spectrum through control of the pump laser. Here, we investigate how $\alpha$ and  $P_{\textrm{in}}$ affect DWs and the DW switching phenomenon. According to Eqn. \ref{0th},  $\alpha$ determines the attenuation rate of $P_{\mu}$ versus $\mu$. In addition, we found both in experiment and simulation that DWs might switch from one mode to another nearby mode during the detuning sweep, enabling fine tuning of DW frequencies. This behavior comes from the nonlinear self-interactions between DWs and its parent soliton, causing $f_{\text{rep}}$ to diverge from FSR \cite{Skryabin2017Self-lockingDispersions}, namely $\lambda\neq0$, which can be regarded as an effective shift in $D_{\textrm{int}}$ and switches DW modes. In Fig. \ref{Fig:pump} (a), we plot three LLE-simulated spectra with varied  $\alpha$: 19 (blue), 26 (green) and 33 (red) ($F^2 = 37.2$). The simulated comb power $P_{\text{comb}}$ versus $\alpha$ is shown in Fig. \ref{Fig:pump} (b), where the corresponding $\alpha$ for each spectrum in Fig. \ref{Fig:pump} (a) is indicated  by diamonds in the same color.  Each comb is a  stable solution of LLE and we have experimentally verified that these states correspond to low noise combs \cite{Briles2018InterlockingSynthesis}. The relative mode numbers of DWs ($\mu_{\text{L}}$, $\mu_{\text{S}}$) switch from (-51,102) (blue) to (-52,100) (green), and then to (-53,99) (red). In contrast to quadratic GVD combs, $P_{\text{comb}}$ does not increase smoothly in Fig. \ref{Fig:pump} (b).  Instead we observe that $P_{\text{comb}}$ vs $\alpha$ behavior is interrupted by high noise regions around $\alpha = 25$ and $32$. These regions are related to the intracavity dynamics of DWs that cause jumps of DW frequency and power \cite{Skryabin2017Self-lockingDispersions}. 

As mention above, $P_{\textrm{in}}$ can also affect CE${}_{\mu}$ by affecting the range of $\alpha$ where the soliton can exist. Figure \ref{Fig:pump} (c) plots three measured spectra from the same device but with increasng $P_{\textrm{in}}$: 74 mW, 89 mW, 126 mW. For each $P_{\textrm{in}}$, we tuned $\alpha$ to maximize the SWDW power in experiment. While $P_{\textrm{in}}$ increases by less than two times, the SWDW power increases by tens of dB, which we attribute to a larger $\alpha$ able to support the soliton.

In summary, microresonator-based optical frequency combs are promising for chip-scale integration and low-power operation but suffer from the high sensitivity to the fabrication process. Benefiting from the consistent fabrication tolerance of a commercial foundry, we demonstrate a systematical design process and wafer-scale search method to achieve octave-spanning combs with an electrically detectable $f_{\textrm{ceo}}$ (5.1 GHz). We also provide an analytical model to predict the soliton spectrum with two DWs and investigate its dependence on the device parameters as well as the pump laser. The experiment results agree well with this analytical model and LLE simulations. Our work represents an important step towards the practical applications of the microresonator-based octave-spanning combs for self referencing.

\noindent \textbf{Acknowledgement.} We thank Atasi Dan for taking the microscope photos. This research has been funded by the AFOSR FA9550-20-1-0004 Project Number 19RT1019, DARPA DODOS, NSF Quantum Leap Challenge Institute Award OMA – 2016244, and NIST. This work is a contribution of the U.S. government and is not subject to copyright. Tradenames provide information only and not an endorsement.

\noindent \textbf{Disclosures.} The authors declare no conflicts of interest.

%\vspace{-5pt} %JZ: we can remvoe this if OL complains about it. 

%\bigskip
%\noindent Add citations manually or use BibTeX. See %\cite{Zhang:14,OSA,FORSTER2007,testthesis,manga_rao_single_2007}.

% Bibliography
\bibliography{sample,references}
%\bibliography{references}

% Full bibliography added automatically for Optics Letters submissions; the following line will simply be ignored if submitting to other journals.
% Note that this extra page will not count against page length
%\bibliographyfullrefs{sample}

%Manual citation list
%\begin{thebibliography}{1}
%\bibitem{Zhang:14}
%Y.~Zhang, S.~Qiao, L.~Sun, Q.~W. Shi, W.~Huang, %L.~Li, and Z.~Yang,
 % \enquote{Photoinduced active terahertz metamaterials with nanostructured
  %vanadium dioxide film deposited by sol-gel method,} Opt. Express \textbf{22},
  %11070--11078 (2014).
%\end{thebibliography}

\clearpage 

%\subsection{Possible References/ideas to add context for introduction}
%\begin{enumerate}
%\item heaters for fceo tuning: \cite{Moille2022IntegratedCombs, Xue2016ThermalResonators}
%\item post fabrication trimming: \cite{Moille2021TailoringDispersion}
%\item pulley couplers: \cite{Moille2019BroadbandMicrocombs}
%\item wafer scale: \cite{Liu2021High-yieldCircuits} (FYI, these figures are worth looking at)
%\item Hybrid integration/interposers: \cite{Rao2021TowardsCombs}
%\item stabilization of microcombs: \cite{Briles2018InterlockingSynthesis,Yu2019TuningResonances}
%\item general freq combs: \cite{Diddams2020OpticalSpectrum}.  \cite{Fortier201920Applications} is another recent review but the published errata is very long...
%\item injection locking????
%\item PhCR to initiate comb
%\end{enumerate}
%%%%% Toggle data details on and off.   %%%%%%
%\include{dataDETAILS.tex}

\end{document}